\documentclass[11pt]{article}

\usepackage[T1]{fontenc}
\usepackage[utf8]{inputenc} 
\usepackage{lmodern}
\usepackage{amsmath,amssymb}
\usepackage{graphicx}
\usepackage{authblk}
\usepackage{moreverb}
\usepackage{hyperref}
\usepackage{fancyhdr}
\usepackage[authoryear]{natbib}
\usepackage{caption}
\usepackage[a4paper,margin=1in]{geometry}
\hypersetup{
  colorlinks=true,
  citecolor=red,
  urlcolor=red,
  linkcolor=red
}

\newcommand{\Da}{\mathrm{Da}}
\newcommand{\F}{\mathrm{Fo}}
\newcommand\BibTeX{{\rmfamily B\kern-.05em \textsc{i\kern-.025em b}\kern-.08em
T\kern-.1667em\lower.7ex\hbox{E}\kern-.125emX}}

\title{Retreat to advance: self-blocking enables efficient mineral replacement}

\author[1]{Agnieszka Budek}
\author[2]{Tomasz Szawe{\l}{\l}o}
\author[3]{Vaughan Voller}
\author[2,*]{Piotr Szymczak}
\affil[1]{Department of Earth and Environmental Sciences, University of Minnesota, Minneapolis, MN, USA}
\affil[2]{Institute of Theoretical Physics, Faculty of Physics, University of Warsaw, Warsaw, Poland}
\affil[3]{Department of Civil, Environmental, and Geo-Engineering and Saint Anthony Falls Laboratory, University of Minnesota, Minneapolis, MN, USA}

\date{}

\begin{document}
\maketitle

\begin{abstract}
Mineral replacement reactions under advective flow often suffer from severe spatial inefficiency: dissolution causes the flow to self-focus into a few dominant wormholes that bypass the surrounding matrix, leaving most of the rock unreplaced.
Here we show---through two-dimensional pore-network simulations---that replacement can be effective in two regimes. The first arises when the precipitation rate significantly exceeds the dissolution rate, leading to in situ replacement in which a uniform front of the secondary mineral advances through the matrix. The second, exploratory mode, occurs when the system repeatedly self-blocks and re-routes. In this regime, each channel lives only long enough to deliver reactant a short distance ahead of the front before its tip is cemented by the product phase; pressure re-routes through an adjacent corridor, and the cycle begins anew. Over time the replacement front advances as a mosaic of overlapping micro-fronts, distributing the secondary mineral almost uniformly. We derive design criteria for achieving exploratory-mode behaviour and discuss implications for both natural and engineered reactive-infiltration systems.
\end{abstract}

\fancypagestyle{plain}{%
  \fancyhf{}
  \renewcommand{\headrulewidth}{0pt}
  \renewcommand{\footrulewidth}{0pt}
  \cfoot{\parbox{0.95\textwidth}{\centering
    \footnotesize
    Accepted manuscript\\
    Published in \emph{G\'eotechnique Letters} (2026). DOI: 10.1680/jgele.25.00094\\
    Licensed under CC BY-NC-ND 4.0 (\url{https://creativecommons.org/licenses/by-nc-nd/4.0/}).
  }}
}
\begingroup
\renewcommand\thefootnote{}
\footnotetext{*Corresponding author: \href{mailto:piotrek@fuw.edu.pl}{piotrek@fuw.edu.pl}}
\endgroup

\section{Introduction}
Coupled dissolution–precipitation reactions lie at the heart of geological transformations~\citep{Korzhinskii1968,Putnis2009,Merino2011,Kondratiuk2017,Beinlich2020,Centrella2021}. Whether converting limestone to dolomite, basalt to zeolite, or serpentinising peridotite, the primary mineral (hereafter A) is eaten away while its ions feed the growth of a secondary phase (E). At first glance, the conversion appears simple: inject a reactive fluid, allow it to dissolve A and let~E precipitate downstream. Yet complete and uniform replacement is often not achieved. Instead, several contrasting behaviours might emerge. At one extreme, if dissolution is much faster than precipitation or if the product occupies less volume than the mineral it replaces,  the flow self‑focuses into a number of pronounced channels (so-called wormholes)~\citep{Daccord1987,Hoefner1988,Szymczak2009} that punch through the rock, bypassing most of the matrix. At the other extreme, if the product is too voluminous, the newly formed mineral chokes the flow paths, causing global clogging and halting the reaction front~\citep{Wagner2005,Yoo2013,Neil2024}.

However, with the right selection of parameters, a regime exists in which replacement proceeds efficiently. This occurs when the precipitation rate is sufficiently high relative to the dissolution rate, such that precipitation takes place in situ,  right at the dissolution front~\citep{Merino1998,Putnis2009}. A second prerequisite is that the molar volume of the secondary phase must be greater than that of the primary phase. Replacement is then associated with a reduction in permeability. This acts as a double-edged sword. It is beneficial because it suppresses reactive-infiltration instability, thereby preventing the formation of wormholes and allowing the precipitation front to advance uniformly. However, if the volume change is too large, clogging can occur. Consequently, in situ replacement is limited to a relatively narrow region in the parameter space characterising dissolution–precipitation reactions.

Here we argue that there is another sweet spot where replacement can be efficient, but for different reasons. This regime exists in a dynamic middle ground where dissolution and precipitation are delicately mismatched so that flow constantly abandons its own channels. This ‘exploratory mode’ maximises the accessible surface area and spreads precipitate widely---two prerequisites for efficient replacement.

An everyday analogy may help to visualise this dynamics. Picture driving across a dense city when the road directly ahead keeps being closed---first by a street festival, then by emergency repairs, next by a protest march, or perhaps by a delivery truck blocking the lane. Each time you must retreat, weave through side streets and chart a new course. The detours lengthen the trip, but if you later trace your trajectory on the map you realise you have explored a broad swath of the city rather than following a single straight line.

\begin{figure*}
\centering
\includegraphics[width=1.\textwidth]{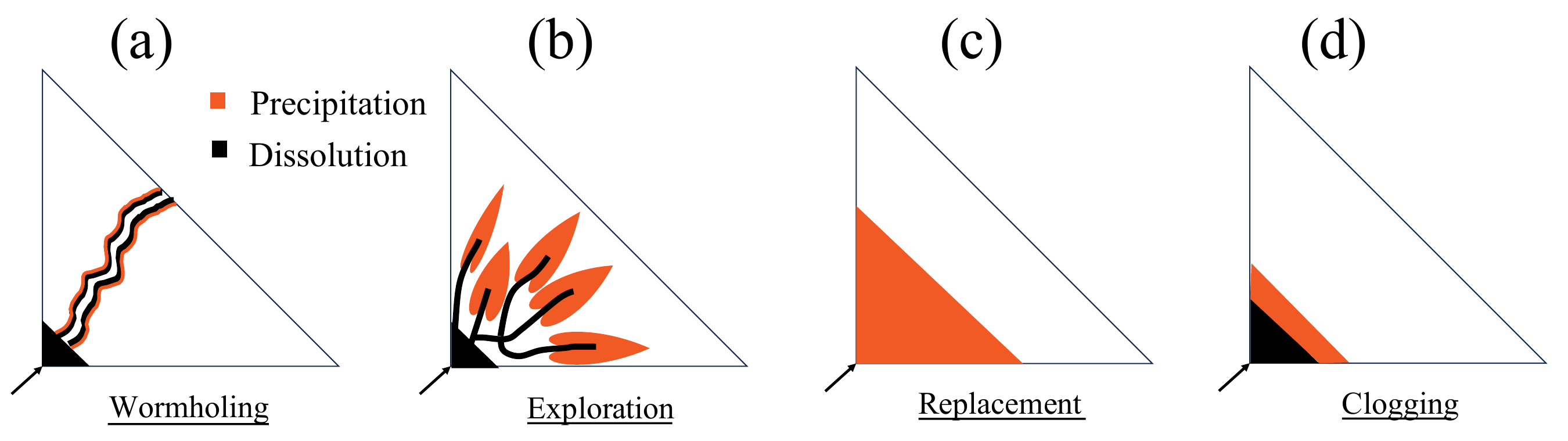}
\caption{Four distinct replacement regimes:
(a) dissolution-dominated: wormholing creates a high-permeability conduit that breaks through to the outlet, confining replacement to the immediate vicinity of the channel;
(b) balanced dissolution–precipitation: wormholes are periodically blocked by precipitate, forcing the flow to re-route and carve new paths;
(c) in situ replacement: precipitation is much faster than dissolution, thus the dissolved material gets immediately precipitated, leading to the in situ replacement of the primary mineral with the secondary phase;
(d) precipitation-dominated: rapid product accumulation completely clogs the network and halts the reaction front.}
\label{fig:cartoon}
\end{figure*}

As illustrated in Fig.~\ref{fig:cartoon}, mineral-replacement fronts in exploratory mode behave similarly: each active flow path is eventually closed by product accumulation, forcing the fluid to backtrack and probe fresh territory, thereby sweeping a far larger volume of rock than a persistent wormhole could ever reach.

Building on~\cite{Budek2025}, who introduced the pore‑network framework for dissolution–precipitation and mapped the space of outcomes, we analyse the emergence of this exploratory regime and how to steer a system into it. In practice, the available levers are the composition and operation of the injected fluid: buffering the precipitating reagent so that precipitation proceeds at a controlled rate relative to dissolution; modest adjustments to pH, ligand chemistry, temperature and background electrolyte that tune kinetics; and flow‑rate or injection‑schedule choices that avoid both early wormholing and premature sealing. 


\section{The model}

To quantify these ideas, we simulate dissolution–precipitation in a two‑dimensional pore network. The medium is represented by cylindrical pores (length $l_{ij}$, diameter $d_{ij}$) connecting nodes; at each step we solve Poiseuille flow and mass conservation,
\[
q_{ij}=\frac{\pi d_{ij}^{4}}{128\,\mu\, l_{ij}}\,(p_i-p_j), \qquad \sum_i q_{ij}=0,
\]
then advance solute in each link with a 1‑D advection–reaction balance for the concentrations of the species, $c^k$,
\begin{equation}
\frac{\partial (q_{ij} c^k)}{\partial x} \;=\; -\,\pi d_{ij} R_k(c^1,\ldots,c^N),
\label{rea}
\end{equation}
where $\pi d_{ij}$ accounts for the reactive surface area  (assumed to be equal to the lateral surface of the pore) and $R_k$ is the
reaction term involving, in principle, the concentrations of all reactive species.
We then solve for the concentrations in the entire network,  
assuming full mixing at junctions~\citep{Park2001,Sharma2023}.

The reactive fluid enters from the upper edge, where pressure $p_{\mathrm{in}}$ is applied, and exits from the lower edge, where the pressure is $p_{\mathrm{out}}=0$. We impose a constant total flow through the system; accordingly, $p_{\mathrm{in}}$ is rescaled at each step. For solute transport, we prescribe Dirichlet inlet concentrations: at the top $c_{\text{B}}=c_{\text{B}}^{\mathrm{in}}$ and $c_{\text{C}}=0$. Periodic boundary conditions are applied along the lateral (left–right) direction.

The system is characterised by the initial pore diameter $d_0$ and average pore length $l_0$, which together set the initial porosity $\varphi_0$, controlling the tendency of pores to clog. 
Heterogeneity arises from variability in the initial pore positions and their connectivity. We do not impose additional variability in pore diameters; although different levels of such variability are expected to shift the boundaries between the observed regimes, they should not alter the qualitative behaviour~\citep{szawello2024}. Note that in our framework, the lattice network is not merely a numerical discretisation of a continuum model but a stylised representation of pore connectivity and structure. 

We consider the simplest dissolution–precipitation reaction, in which the dissolution of mineral A by reactant B is coupled by a common ion C with precipitation of mineral E,
\begin{equation}
\chi_{\text{A}} \text{A} + \chi_{\text{B}} \text{B} \rightarrow \text{C}^\uparrow,
\label{e1}
\end{equation}
\begin{equation}
\text{C} + \chi_{\text{D}} \text{D} \rightarrow \chi_{\text{E}} \text{E}^\downarrow,
\label{e2}
\end{equation}
where $\chi_k$ are the stoichiometric coefficients for the reactions. 
A similar dissolution–precipitation coupling through a common ion is at the base of mineral carbonation processes 
\citep{Andreani2009,Matter2009,Sanna2014}. One of the challenges here is to assess how to prevent self-clogging of this reaction under the conditions when the precipitated product (E) has an equal or larger molar volume than the dissolved mineral (A), as is the case in most of the mineral carbonation reactions.

We assume that dissolution reaction~(Equation 2) is irreversible and first order in the concentration of the dissolving species $c_{\text{B}}$,
\begin{equation}
R_{\mathrm{diss}} = k_{\mathrm{diss}}\, c_{\text{B}},
\label{rates}
\end{equation}
where $k_{\mathrm{diss}}$ is the dissolution rate constant. 
Precipitation~(Equation 3) is taken to be second order in the concentrations of aqueous reactants $c_{\text{C}}$ and $c_{\text{D}}$,
\begin{equation}
R_{\mathrm{prec}} = k'\, c_{\text{C}}\, c_{\text{D}} ,
\end{equation}
where $k'$ is the respective rate constant. However, 
we further assume that D is in large excess or buffered, so that its concentration is approximately constant, $c_{\text{D}} \approx c_{\text{D}}^\ast$. In this case the precipitation rate reduces to a pseudo–first-order form,
\begin{equation}
R_{\mathrm{prec}} = k_{\mathrm{prec}}\, c_{\text{C}}, 
\qquad k_{\mathrm{prec}} := k' c_{\text{D}}^\ast,
\label{rate2}
\end{equation}
with an effective precipitation rate constant $k_{\mathrm{prec}}$. These rate laws are appropriate when the system is far from equilibrium and rates are surface-reaction-limited, with background composition effectively constant~\citep{lasaga1998}.
More complex kinetics can be incorporated if warranted by data; here we adopt the simplest forms that capture the regimes of interest and keep the presentation focused.

\begin{figure*}
\centering
\includegraphics[width=0.83\textwidth]{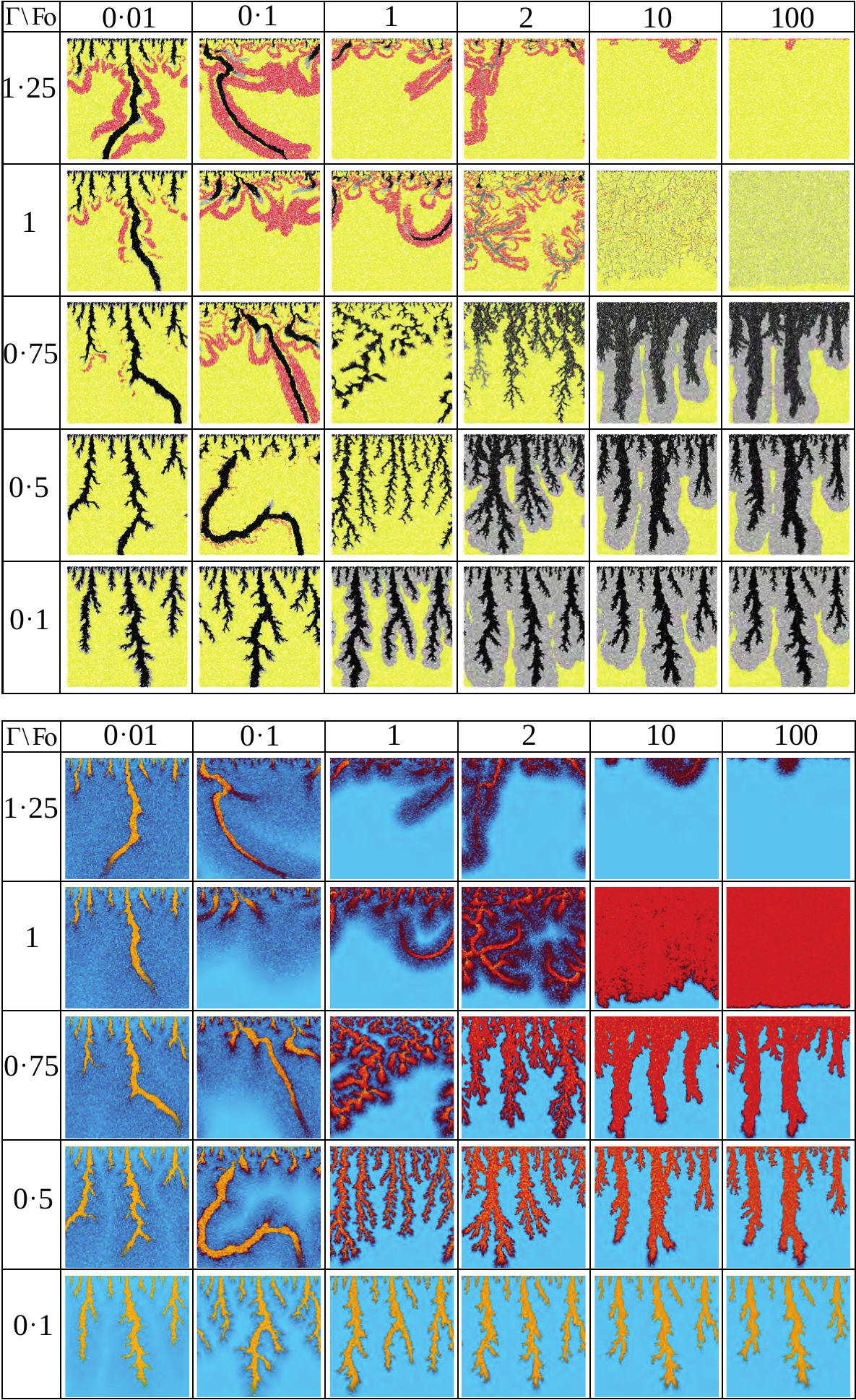}
\caption{Morphological phase diagram of dissolution-precipitation patterns as a function of $\F$ and $\Gamma$ ($\Da=0.5$, $\varphi_0=0.04$).
The simulations are carried out on a grid of $200\times 200$ randomly distributed nodes, continuing until either a continuous fluid pathway develops (breakthrough) or the system becomes completely sealed by mineral precipitation (clogging). Upper panel: pore diameter maps. Pores that are heavily overgrown, $d \leq d_0/10$, are marked in red; intermediate pores, $d_0/10 < d \leq d_0$, are yellow; pores larger than $d_0$, $d_0 < d \leq 2d_0$, are grey; pores creating dissolution pattern, $d > 2d_0$, are marked in black. Lower panel: mineral content maps. Colours indicate mineral compositions of the grains: mineral A is shown in blue, mineral E in red, and mixtures of the two appear in shades of purple; pore space (not occupied by grains) is yellow. Adapted from~\citet{Budek2025}.}
\label{f2}
\end{figure*}
Given the reaction rates, the change of radii of the pores can be calculated based on the mass conservation
\begin{equation}
\partial_t (d_{ij}/2) = R_{\text{diss}} \chi_{\text{A}} \nu_{\text{A}} - R_{\text{prec}} \chi _{\text{E}} \nu_{\text{E}}  ,
\end{equation}
where $\nu_{\text{A}}$ and $\nu_{\text{E}}$  are the molar volumes of the dissolved and precipitated rock material A and E respectively.
Thus, the local expansion of each pore element is directly proportional to dissolution and/or precipitation rate at each point. To keep the system mathematically tractable, an approximation is usually adopted in which the diameter of each pore element is assumed to be constant along its length. Consequently, the reactive flux is integrated along the pore, and the resulting dissolved/precipitated volume is uniformly distributed along its length \citep{Hoefner1988,Budek2012}. 

Because flow and transport relax much faster than geometry, we use a quasi-static update (fields $q$ and $c$ are steady within each step)~\citep{Lichtner1988}. With a constant inflow $Q_{\mathrm{in}}$, we measure time in injected pore volumes (PV):
\begin{equation}
\mathrm{PV}=\frac{Q_{\mathrm{in}}\, t}{V_{\mathrm{pore}}^0},
\label{eq:pv_constQ_simple}
\end{equation}
where $V_{\mathrm{pore}}^0$ is the initial pore volume. In practice we advance geometry with small steps $\Delta t$ and accumulate $\Delta\mathrm{PV}=Q_{\mathrm{in}}\Delta t/V_{\mathrm{pore}}^0$. All time axes are reported in PV.

There are three important dimensionless parameters governing this process. The first is the Damk\"ohler number:
\begin{equation}
\Da = \frac{\pi k_{\text{diss}} d_0 l_0}{q_0},
\end{equation}
where $q_0$ denotes the initial average volumetric flow rate within the pores. 
The second dimensionless parameter is the ratio of precipitation to dissolution rate constants:
\begin{equation}
\F = \frac{k_{\text{prec}}}{k_{\text{diss}}}.
\end{equation}
We propose naming it the Fogler number, after H.~Scott Fogler, in recognition of his seminal contributions to understanding pattern formation in rocks~\citep{Hoefner1988,Fredd1998}, particularly his pioneering work elucidating the interplay between dissolution and precipitation~\citep{Rege1989}.

Finally, the third parameter is the ratio of the molar volume of the precipitating mineral to that of the dissolving mineral:
\begin{equation}
\Gamma = \frac{\chi_{\text{E}} \nu_{\text{E}}}{\chi_{\text{A}} \nu_{\text{A}}}.
\end{equation}
This ratio plays a crucial role in the evolution of the geometry: $\Gamma<1$ corresponds to reactions that create porosity, whereas $\Gamma>1$ yields a net increase in solid volume, reducing porosity and potentially causing self-clogging.

\section{Results}

Fig.~\ref{f2} reveals how dissolution–precipitation patterns evolve under varying conditions defined by the parameters $\F$ and $\Gamma$. At the lowest precipitation rate ($\F = 0.01$), the common ion C is advected out of the network faster than it can precipitate, so clogging does not occur. Here, small precipitate molar volume ($\Gamma \ll 1$) yields patterns resembling classical dissolution wormholes. As $\Gamma$ increases, precipitate accumulates around dissolution channels, acting as barriers that inhibit branching and guide wormholes into straighter forms.
At a moderate precipitation rate ($\F=0.1$), precipitate forms closer to the active dissolution front, creating cemented zones---visible in the Figure as intense red---that build walls around the advancing pathways and steer the channel's further course.  Eventually, large values of $\Gamma$ lead to the clogging of the system. 

At higher $\F$, precipitation occurs essentially at the front, reducing the likelihood of clogging---except at large $\Gamma$---and promotes the formation of straighter, less-branched channels. As $\Gamma$ increases, the permeability contrast between these channels and their surroundings diminishes; they then coexist, grow thicker and sometimes form loops ($\F \geq 10$, $\Gamma = 0.75$). When $\Gamma \geq 1$, we enter the in situ replacement regime: the permeability contrast is eliminated entirely, halting instability and producing a uniform, non-branching replacement front, unless $\Gamma$ becomes sufficiently large to induce clogging.

The richest morphological variety emerges at intermediate $\F$ and $\Gamma$ (e.g., $\F = 1$, $\Gamma = 0.75$; or $\F=2$, $\Gamma = 1$). This is the exploratory mode, in which intense interaction between dissolution and precipitation generate intricate, multi-path networks by way of recurrent cycles of partial clogging and reopening. During an individual cycle, precipitation locally reduces porosity and increases hydraulic resistance, diverting flow towards neighbouring pores; the diverted flux enhances dissolution there, re-establishing throughput until precipitation again accumulates and the process migrates. These dynamics yield large, quasi-periodic fluctuations in permeability.  

\begin{figure*}
\centering
\includegraphics[width=1.\textwidth]{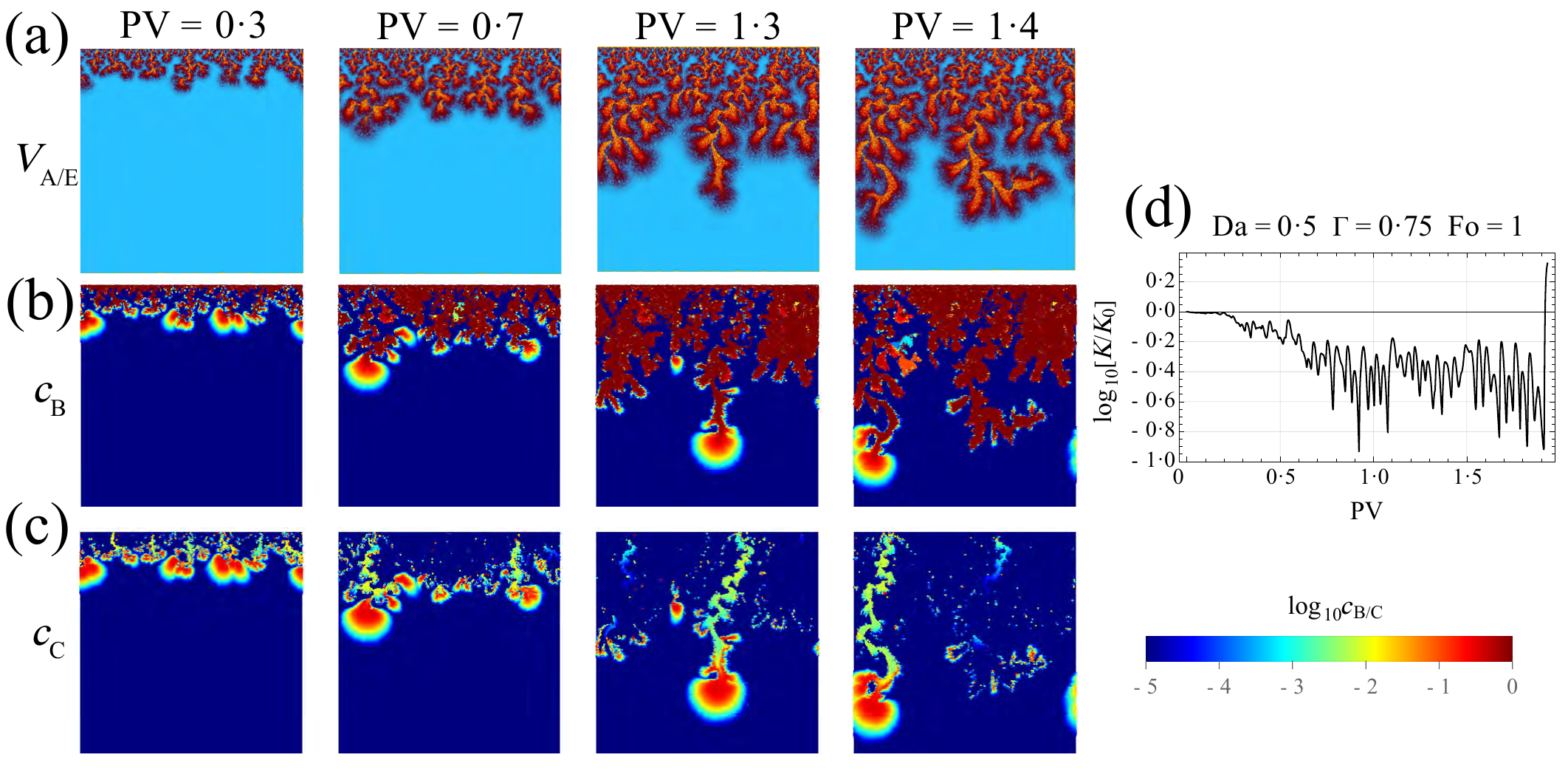}
\caption{(a–c): Snapshots of system evolution in the exploratory regime ($\F=1, \Gamma = 0.75, \Da=0.5, \varphi_0=0.04$). The simulation uses a network of $400\times 400$ randomly distributed nodes. (a) Mineral map: mineral A (blue), mineral E (red), mixtures (purple); pore space (yellow). (b–c) Concentration fields of species
B and C, respectively. Only a small subset of channels conveys solute deeper into the system. (d) 
Relative permeability $K/K_0$, as a function of time (in injected pore volumes)  for the same run.}
\label{fig:oscillations}
\end{figure*}

Figure~\ref{fig:oscillations} illustrates these cycles: panel (a) shows the advancement of the replacement pattern; panels (b) and (c) track the corresponding spatiotemporal shifts of reactant concentration, as throughput relocates when existing pathways become obstructed and new ones emerge. In space, the pattern remains broadly distributed rather than collapsing into a single wormhole; in time, the system explores alternative routes, maintaining multiple partially active paths. The net effect is efficient, system-wide dispersal of the secondary phase, accompanied by large oscillations in the effective permeability, as shown in panel (d).

\begin{figure*}[h]
\centering
\includegraphics[width=1.\textwidth]{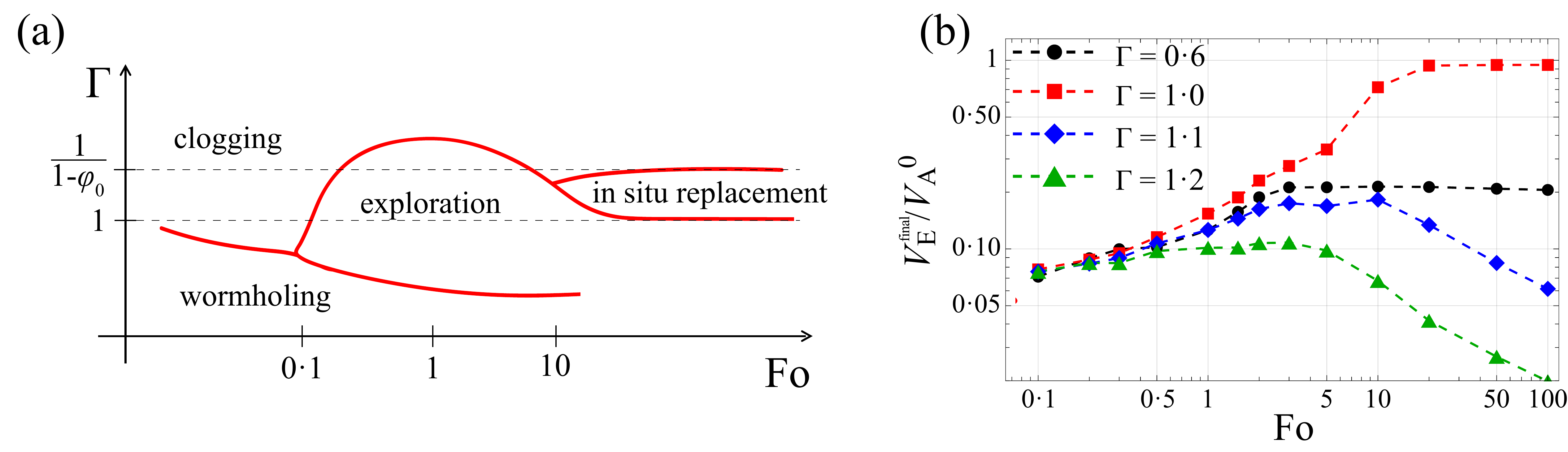}
\caption{(a) Schematic morphologies in $\F$–$\Gamma$ space. The regime boundaries are approximate; their locations depend on the initial porosity $ \varphi_0$, the Damk\"ohler number $\Da$ and the size of the system. (b) Volume of the precipitated mineral $V_{\text{E}}^{\text{final}}$, normalised by the initial volume of the primary mineral $V_{\text{A}}^0$, as a function of $\F$ for $\Da = 0.5$, $\varphi_0 = 0.04$ and $\Gamma=0.6$, $1.0$, $1.1$ and $1.2$. The statistics are averaged over $40$ realizations on $100\times100$ random-node grids, with each simulation terminated at either clogging or breakthrough.}
\label{f3}
\end{figure*}

In Fig.~\ref{f3}(a), replacement regimes are mapped across the parameter space. At low $\F$ and small $\Gamma$, dissolution dominates and wormholing occurs. At high $\F$ with $\Gamma\!\gtrsim\!1$, tip instabilities are suppressed and in situ replacement advances as a near-uniform front. For intermediate $\F$ and $\Gamma$, an exploratory regime emerges: repeated self-blocking and rerouting maintain multiple active paths and broadly distribute the secondary phase. For sufficiently large $\Gamma$, precipitation increases solid volume faster than dissolution creates void space, so flow paths close and the network clogs.

Figure~\ref{f3}(b) quantifies replacement efficiency as a function of $\F$; the trends depend strongly on $\Gamma$. Peak efficiency appears near $\Gamma=1$ at large $\F$---the in situ replacement regime. Although this regime produces the largest volume of precipitate, it spans only a narrow range of $\Gamma$: the lower bound is $\Gamma = 1$, required to keep the front stable, while the upper bound, $\Gamma = \tfrac{1}{1-\varphi_0}$, marks the point at which the secondary phase completely replaces the primary mineral while filling the entire available pore space, driving the porosity to zero~\citep{Budek2025}. Once $\Gamma$ exceeds that limit, high-$\F$ efficiency collapses and the exploratory regime at intermediate $\F$ becomes the most effective. At small $\Gamma$, the efficiency depends much less on $\F$ because wormholing dominates the process.

\section{Experiments}
Bench-scale experiments document the same competition between channel opening and self-blocking that we emphasise here. \citet{Rege1989} injected acid/iron solutions into carbonate cores and, by varying the inlet concentrations, effectively sampled different regions of the morphological diagram. In the mixed dissolution-precipitation regime they observed oscillations in permeability consistent with alternating phases of channel growth and clogging. They also cast the final pore structures by infiltrating Wood’s metal, showing that dissolution–precipitation interplay yields markedly more ramified patterns that engage a larger fraction of the pore space than pure‑dissolution runs at the same flow rate.

In carbonate sandstone columns,~\citet{singurindy2003a} flooded calcareous sandstone with an HCl/H$_2$SO$_4$ mixture, driving calcite dissolution coupled to gypsum precipitation (here sulfate plays the role of species D in our notation; $\Gamma \approx 2$ for calcite $\rightarrow$ gypsum). At high $[\mathrm{H}^+]/[\mathrm{SO}_4^{2-}]$ ratios (low sulfate, hence small $k_{\mathrm{prec}}$ and low $\F$), dissolution dominated and wormholes formed; at low $[\mathrm{H}^+]/[\mathrm{SO}_4^{2-}]$ ratios (high sulfate, large $k_{\mathrm{prec}}$ and $\F$), precipitation dominated and the columns clogged. At intermediate acid–sulfate ratios, they reported strong spatial porosity variations: active dissolution channels (wormholes) juxtaposed with gypsum‑cemented zones, together with oscillatory permeability during the runs.

As shown in Fig.~\ref{fig:oscillations}, oscillations in permeability also occur in our system and are a direct signature of the exploratory regime. The opening of a new pathway is associated with an increase in permeability, followed shortly by a decrease as that pathway closes. These oscillations provide a practical diagnostic of exploratory behaviour and can serve as feedback to steer experiments into this regime. In experimental systems, however, additional complexity can arise because D is typically unbuffered, introducing extra feedback between precipitation rate and transport. We will examine this feedback in future work.

\section{Discussion}
The morphological diagram spanned by $(\F,\Gamma)$ admits direct experimental control. In particular, because the precipitation kinetics obey $k_{\mathrm{prec}}=k' c_{\text{D}}^\ast$ under buffered conditions, tuning the buffered level (or supply rate) of species D directly adjusts $k_{\mathrm{prec}}$ and therefore shifts the system’s effective precipitation-to-dissolution propensity, $\F$. By increasing or decreasing $c_{\text{D}}^\ast$, one can deliberately navigate across the diagram towards the ‘sweet spot’ between dissolution- and precipitation-dominated regimes, where channel competition remains active yet clogging is delayed. In this intermediate regime, the channels explore a much larger fraction of the domain before a single pathway takes over or the network cements, which translates into markedly higher replacement efficiency. 
When $\Gamma \approx 1$, increasing $\F$ drives a transition from the exploratory regime to in situ replacement---a steadier and more spatially uniform mode. In practice, however, the in situ pathway operates only over a narrow band $\Gamma$, whereas the exploratory regime remains effective across a much broader span of $\Gamma$ values. Consequently, the exploratory regime is the more robust option under varied geochemical conditions.

We note that in situ replacement with $\Gamma > 1$ has also been associated in the literature with reaction-driven cracking, where volume expansion generates crystallisation stresses that fracture the host rock. This process provides another pathway for sustaining permeability in otherwise sealing systems \citep{kelemen2012,feng2025}.

Practically, these trends suggest a clear protocol for targeting high-efficiency replacement: choose mineral pairs or solution chemistries that yield a moderate $\Gamma$ and use $k_{\mathrm{prec}}$ (through buffer strength, background electrolyte composition or controlled dosing of D) to place the system at intermediate $\F$. Operating in this band maximises spatial exploration by sustaining multiple active channels while avoiding premature cementation, thereby converting substantial regions of the domain before either a breakthrough concentrates the flow into a single wormhole or widespread clogging halts progress.

In summary, the interplay of flow, transport and reaction in such a system can give rise to a variety of patterns, from spontaneous channelling to nearly homogeneous transformation of the entire rock matrix into the product phase. Interestingly, even if the product phase has a larger molar volume than the primary phase, clogging in such a system can be avoided, due to the interplay of dissolution and precipitation resulting in the continuous creation of new flow paths. The exploratory regime identified here is directly actionable in both natural and engineered settings. In the latter, tuning the buffered species that sets $k_{\mathrm{prec}}$ (and thus $\F$) provides a handle to operate near the exploration optimum, suppressing premature sealing while promoting uniformly distributed precipitation of the secondary phase. In hydrothermal replacement systems, intermittent self‑blocking yields laterally migrating micro‑fronts and looped conduits, a mechanism for broad, efficient replacement over substantial areas when $\F$ and $\Gamma$ are near the efficiency peak. In reservoir stimulation and geothermal acidising, introducing a controlled, fast‑but‑buffered precipitating pathway can prevent single‑wormhole dominance, maintain multiple active paths and improve sweep and heat exchange.

\section*{Data availability}
The datasets generated and/or analysed during the current study are available from the corresponding author on request. 

\section*{Acknowledgements}
This work was supported by the National Science Centre (NCN, Poland) under Grant No. 2022/47/B/ST3/03395. We thank Peter K. Kang for helpful discussions. Piotr Szymczak acknowledges the support through the Fulbright STEM Award programme for funding his visit at the University of Minnesota.
Vaughan Voller was supported by the  Center on Geo-processes in Mineral Carbon Storage, an Energy Frontier Research Center funded by the U.S. Department of Energy, Office of Science, Basic Energy Sciences at the University of Minnesota under Award No. DE-SC0023429.

\bibliographystyle{plainnat}

\end{document}